\documentclass[aps,prb,twocolumn,showpacs,superscriptaddress]{revtex4-1}
\usepackage{amsmath}
\usepackage{graphicx}
\usepackage{subfigure}
\usepackage{color}
\begin{document}
\title{Anatomy of electrical signals and dc-voltage lineshape in spin
torque ferromagnetic resonance}
\author{Yin Zhang}
\affiliation{Physics Department, The Hong Kong University of Science and
Technology, Clear Water Bay, Kowloon, Hong Kong}
\affiliation{HKUST Shenzhen Research Institute, Shenzhen 518057, China}
\author{Q. Liu}
\affiliation{National Laboratory of Solid State Microstructures and
Department of Physics, Nanjing University, 22 Hankou Road, Nanjing 210093, China}
\author{B. F. Miao}
\author{H. F. Ding}
\email{[Corresponding author:]hfding@nju.edu.cn}
\affiliation{National Laboratory of Solid State Microstructures and
Department of Physics, Nanjing University, 22 Hankou Road, Nanjing 210093, China}
\affiliation{Collaborative Innovation Center of Advanced Microstructures,
Nanjing University, 22 Hankou Road, Nanjing 210093, China}
\author{X. R. Wang}
\email{[Corresponding author:]phxwan@ust.hk}
\affiliation{Physics Department, The Hong Kong University of Science and
Technology, Clear Water Bay, Kowloon, Hong Kong}
\affiliation{HKUST Shenzhen Research Institute, Shenzhen 518057, China}
\date{\today}

\begin{abstract}
The electrical detection of spin torque ferromagnetic resonance (st-FMR) is
becoming a popular method for measuring the spin-Hall angle of heavy metals (HM).
However, various sensible analysis on the same material with either the same or
different experimental setups yielded different spin-Hall angles with large discrepancy,
indicating some missing ingredients in our current understanding of st-FMR.
Here we carry out a careful analysis of electrical signals of the st-FMR
in a HM/ferromagnet (HM/FM) bilayer with an arbitrary magnetic anisotropy.
The FM magnetization is driven by two radio-frequency (rf) forces:
the rf Oersted field generated by an applied rf electric current
and the so called rf spin-orbit torque from the spin current
flowing perpendicularly from the HM to the FM due to the spin-Hall effect.
By using the universal form of the dynamic susceptibility matrix of magnetic
materials at the st-FMR, the electrical signals originated from the
anisotropic magnetoresistance, anomalous Hall effect and inverse spin-Hall
effect are analysed and dc-voltage lineshape near the st-FMR are obtained.
Angle-dependence of dc-voltage is given for two setups.
A way of experimentally extracting the spin-Hall angle of a HM is proposed.
\end{abstract}
\maketitle

\section{Introduction}

Ferromagnetic resonance (FMR) is a traditional method for extracting magnetic
material parameters such as magnetization, magnetic anisotropy and damping
coefficient \cite{kittel,polder,suhl,tannenwald,seidel,lax,farle,schlomann,
ignatchenko,celinski,boris,vittoria,counil} by either measuring microwave
absorption or detecting electrical signals \cite{juretschke,egan,saitoh,
costache,hoffmann,buhrman,azevedo,nakayama,kondou,hfding,ohno,ysgui,
harder,tulapurkar,sankey,goennenwein,yamaguchi,wang2017,xdtao}.
The microwave absorption spectroscopy is the first generation of FMR technique.
It typically requires large samples in order to have detectable absorption signal.
The analysis is relatively simple because it uses the field-dependence of FMR
peak and the peak width to probe the magnetization and damping. In the electrical
detection of FMR, sample sizes can be very small due to the high electrical signal
detection. Its analysis is, however, more involving although electrical detection
can be at very high precision and samples have less effect on microwave fields.
The electrical signals can come from the anisotropic magnetoresistance (AMR),
anomalous Hall effect (AHE) \cite{juretschke,egan,harder}, as well as the
recently discovered inverse spin-Hall effect (ISHE) \cite{mizukami,mzwu,
hirsch,bauer,tserkovnyak,manchon}. This technique has been widely used in
recent years to extract the spin-Hall angle of heavy metals that measures the
spin-charge interconversion efficiency in both the spin-Hall effect (SHE) and
ISHE \cite{saitoh,costache,hoffmann,buhrman,azevedo,nakayama,kondou,hfding,ohno}.
The spin Hall angle of a heavy metal (HM) is typically measured from the
HM/ferromagnet (HM/FM) bilayers. The FM can be a metal or an insulator.
The FMR is triggered by a microwave in cavity or coplanar waveguide \cite
{saitoh,hoffmann,azevedo,nakayama,hfding,ohno,ysgui,harder}.
The typical setup in an FMR is to eliminate effect of the microwave electric
field on magnetization dynamics so that microwave magnetic field is assumed
to be the only driving force of the FMR. So far, the experimentally
extracted values show a large discrepancy for the same materials even with
similar experimental setups. For example, the measured spin Hall angle of
Pt varies from 0.013 to 0.08 \cite{hoffmann,buhrman,azevedo,kondou,xdtao}.
This large discrepancy comes from many different sources although it is often
attributed to the inaccuracy in mixing conductance of HM/FM interface and spin
diffusion length of the HM. For example, the dynamic susceptibility at the FMR
is in general a non-Polder tensor \cite{wang2017} that depends on the magnetic
anisotropy and damping constant, but it is commonly treated as scalar numbers
or at most a Polder tensor in experimental analysis. Also, the electrical
signal is very sensitive to the phase difference between rf magnetic and
electric fields inside a sample \cite{azevedo,hfding,harder,wang2017}.
This phase difference is not easy to determine accurately in experiments.
In general, the analysis for both HM/FM-metal and HM/FM-insulator are complicated.
For metallic FM, one needs to separate the contribution of ISHE from those
of AMR and AHE through a very careful analysis in an experimental setup
\cite{saitoh,costache,hoffmann,buhrman,azevedo,hfding,ohno,wang2017}.
Although there is no electrical signal in an insulator so that no AMR and AHE
contributions to dc-voltage from the FM insulator, the amount of spin current pumped
from FM through the HM/FM interface is an issue, in particular when new effects
like the spin-Hall magnetoresistance is considered \cite{bfmiao,Chiba,Schreier}.

In recent years, the spin torque ferromagnetic resonance (st-FMR) is becoming
another popular method for measuring spin-Hall angle where an rf current is
directly applied in the sample \cite{buhrman,kondou}. In this technique, there
are two driving forces. One is rf Oersted field generated from rf current applied
in the bilayer. The other is so-called the rf spin-orbit torque (SOT) from the
spin current flowing perpendicularly from the HM to the FM due to SHE.
The magnetization can resonate with both rf Oersted field and the rf SOT.
Compare with microwave FMR, st-FMR does not have phase difference problem
between rf electric and magnetic fields since the Oersted field
is in-phase with rf current. However, the spin-Hall angle was often
over-estimated \cite{hoffmann,buhrman,azevedo,kondou}, which indicates
some missing ingredients in our current understanding of st-FMR.
Thus, a careful analysis of electrical signals of the st-FMR is timely important.

In this work, we perform an anatomy of electrical signals and dc-voltage
lineshape in st-FMR. The paper is organized as follows.
In section 2, we first describe the model and approach adopted in this study.
By using the universal form of the dynamic susceptibility matrix of magnetic
materials at FMR, we analyze the electrical signals originated from AMR, AHE
and ISHE, and obtain the dc-voltage lineshape near the st-FMR. A recipe for
extracting the spin-Hall angle of the HM from the experiments is proposed.
In section 3, the theoretical angle-dependence of dc-voltage is obtained
for two experimental configurations. In the discussion, based on general
physics principles, we argue possible new SHEs and ISHEs in magnetic materials
when the charges, spins and orbits mutually interact among themselves.
The conclusion is given in section 4, followed by the acknowledgements.

\section{Theoretical analysis}

\subsection{Model and analysis}

\subsubsection{Setup and magnetization dynamics}

\begin{figure}
\centering
\includegraphics[width=0.48\textwidth]{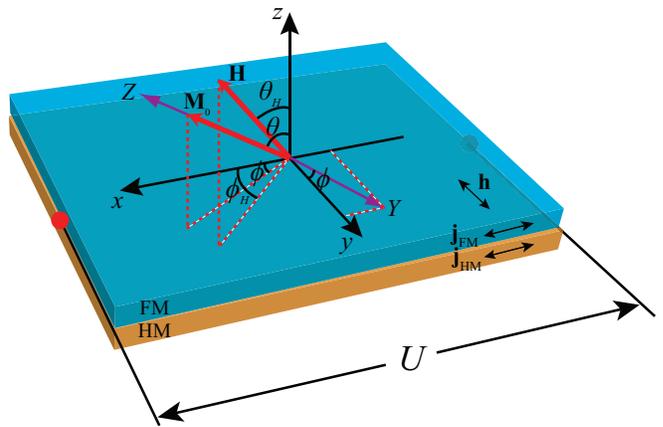}\\
\caption{Model system that mimics the experimental setups of st-FMR.
The $xyz$-coordinate is fixed with respect to the sample.
The HM/FM bilayer sample lies in the $xy$-plane.
The $XYZ$ is a moving coordinate with the $Z$-axis along $\mathbf M_0$,
and the $Y$-axis in the $xy$-plane. $\theta$ and $\phi$ are the polar
and azimuthal angles of $\mathbf M_0$ in the $xyz$-coordinate, i.e.
$\theta$ is the angle between the $Z$- and $z$-axes, and $\phi$ is
the angle between the in-plane component of $\mathbf M_0$ and the
$x$-axis. $\theta_H$ and $\phi_H$ are the polar and azimuthal angles
of the external static magnetic field $\mathbf H$ in the $xyz$-coordinate.
$\mathbf j_\mathrm{FM}$ and $\mathbf j_\mathrm{HM}$ are respectively
the rf electric current in the FM and HM layer.
}
\label{configs_chi}
\end{figure}

The st-FMR model consists of a HM/FM bilayer lying in the $xy$-plane,
as shown in Fig. \ref{configs_chi}. $\mathbf M$ is the magnetization of FM.
A static magnetic field $\mathbf H$ together with an rf current density
$\mathbf J_a=\mathrm{Re}(\mathbf j_a e^{-i\omega t})$ ($a=$FM, HM) is
applied in the bilayer where $\omega$ is the microwave frequency.
Without the rf current, the magnetization is along $\mathbf M_0$. To simplify
the analysis, we use two Cartesian coordinates. The $xyz$-coordinate is fixed
with respect to the sample while the $XYZ$ is a moving coordinate with the
$Z$-axis along $\mathbf M_0$, and the $Y$-axis in the $xy$-plane.
$\theta$ and $\phi$ are the polar and azimuthal angles of $\mathbf M_0$ in
the $xyz$-coordinate, i.e. $\theta$ is the angle between the $Z$- and $z$-axes,
and $\phi$ is the angle between the in-plane component of $\mathbf M_0$ and
the $x$-axis. $\theta_H$ and $\phi_H$ are the polar and azimuthal angles
of the external static magnetic field $\mathbf H$ in the $xyz$-coordinate.
Therefore, once $\mathbf M_0$ is determined, unit vectors $\hat Z$, $\hat X$
and $\hat Y$ are respectively $\hat Z=\sin\theta\cos\phi\hat x+\sin\theta\sin
\phi\hat y+\cos\theta\hat z$, $\hat X=\cos\theta\cos\phi\hat x+\cos\theta\sin\phi
\hat y - \sin\theta\hat z$
and $\hat Y=-\sin\phi\hat x+\cos\phi\hat y$.

Under a microwave radiation, the rf electric current in HM generates an rf
transverse spin current $\mathbf J_s=\mathrm{Re}(\mathbf j_s e^{-i\omega t})$
perpendicularly flowing into the FM layer via the SHE \cite{hirsch} where
the polarization $\mathbf j_s=(\frac{\hbar}{2e})\theta_\mathrm{SH}\mathbf j_\mathrm{HM}\times\hat z$.
Spin Hall angle $\theta_\mathrm{SH}$ measures the conversion efficiency
between charge and spin. The SOT on the magnetization induced by the spin
current is \cite{manchon,Slonczewski,Berger},
\begin{equation}
\vec\tau=-\gamma\frac{a}{M}\mathbf M\times(\mathbf M\times\mathbf J_s)
+\gamma\beta a\mathbf M\times\mathbf J_s,
\label{sot}
\end{equation}
where the first term on the right-hand-side is the Slonczewski-like torque while
the second term is the field-like torque. $a=\frac{1}{d_\mathrm{FM}\mu_0M}\eta$
where $d_\mathrm{FM}$ and $\mu_0$ are respectively the thickness of the FM layer
and the permeability constant.
$\eta$ measures the efficiency of spin angular momentum transfer from the
spin current to the magnetization. $\beta$ measures the
field-like torque and can be an arbitrary real number since this torque
may also be directly generated from the Rashba effect \cite{manchon}.

The magnetization dynamics under a microwave radiation is governed
by the generalized Landau-Liftshitz-Gilbert (LLG) equation \cite{gilbert},
\begin{equation}
\frac{\partial\mathbf M}{\partial t}=-\gamma\mathbf M\times\mathbf H_\mathrm
{eff}+\frac{\alpha}{M}\mathbf M\times\frac{\partial\mathbf M}{\partial t}
+\vec\tau,
\label{llgs_eq}
\end{equation}
where $\gamma$ is the gyromagnetic ratio, $\alpha$ is the Gilbert damping
coefficient, and $\mathbf H_\mathrm{eff}$ is the effective field which includes
the applied static magnetic field $\mathbf H$, rf Oersted field $\mathrm{Re}
(\mathbf h e^{-i\omega t})$ generated by the rf current in the system and
anisotropy field. We assume that the microwave skin depth is much larger than
the FM thickness $d_\mathrm{FM}$, so that the rf current $\mathbf j_\mathrm{FM}$
in the FM layer is spatially uniform and the Oersted field from $\mathbf
j_\mathrm{FM}$ produces no net torque on magnetization. Therefore, the rf Oersted
field is only from $\mathbf j_\mathrm{HM}$. Under the condition that the sample
width is much larger than the HM thickness $d_\mathrm{HM}$, the rf magnetic field
can be determined by the Ampere's law,
i.e., $\mathbf h=\frac{d_\mathrm{HM}}{2}\mathbf j_\mathrm{HM}\times\hat z$.

In the linear response regime, $\mathbf M=\mathbf M_0+\mathrm{Re}(\mathbf me^{-i
\omega t})$ will deviate from its static value $\mathbf M_0$ by a small amount under
the rf Oersted field $\mathbf h$ and rf SOT $\vec\tau$ of frequency of $\omega$.
They are from the same physical origin as rf SOT is originated from the rf
spin current that converted from $j_\mathrm{HM}$ via SHE.
Although the sources of the rf Oersted field $\mathbf h$ and rf SOT are the same,
it is convenient to consider them as two separated forces of magnetization.
Off the resonance, the magnitude of $\mathbf m$ is negligibly small so that no
detectable electrical signal exists. Near the resonance, the responses of $\mathbf m$
to rf field $\mathbf h$ and rf $\mathbf e $ (or rf SOT) are large and are
characterized by the dynamic susceptibilities $\tensor\chi$ and $\tensor\kappa$
defined as $\mathbf m=\tensor\chi\mathbf h+\tensor\kappa\mathbf e$. $\mathbf e$
generates $\mathbf j_\mathrm{HM}$ that is the ultimate source of $\mathbf h$ and rf SOT.
Thus, $\tensor\kappa$ and $\tensor\chi$ are related to each other (see the next subsection).

\subsubsection{Origins of dc-voltage}

In a magnetic field, $\mathbf m$ and $\mathbf j_\mathrm{FM}$, as well as $\mathbf j_
\mathrm{HM}$, are not in phase because $\tensor\kappa$ and $\tensor\chi$ are complex tensors.
Thus $\mathbf j_\mathrm{FM}$ feels an oscillating resistance due to AMR and AHE that has
a phase lag with $\mathbf j_\mathrm{FM}$, resulting in spin rectification effect.
The phase lag also results in a dc-spin-current so that a dc-voltage can also appear
in HM from ISHE. In summary, dc-voltage comes from AMR, AHE, and ISHE,
\begin{equation}
U=U_\mathrm{AMR}+U_\mathrm{AHE}+U_\mathrm{ISHE}.
\end{equation}
According to the generalized Ohm's law \cite{wang2017} in which the AMR and AHE couple
the magnetization motion $\mathbf m$ with the rf electric current $\mathbf j_\mathrm{FM}$,
$U_\mathrm{AMR}$ and $U_\mathrm{AHE}$ are \cite{wang2017}:
\begin{equation}
\begin{aligned}
U_\mathrm{AMR}
=\frac{\Delta\rho}{2M}\mathrm{Re}\{[(\mathbf j_\mathrm{FM}^\ast\cdot\mathbf m)l_Z
+j_{\mathrm{FM},Z}^\ast(\mathbf m\cdot\mathbf l)]\},
\end{aligned}
\label{dc_amr}
\end{equation}
\begin{equation}
\begin{aligned}
U_\mathrm{AHE}
=-\frac{R_1}{2}\mathrm{Re}[(\mathbf j_\mathrm{FM}^\ast\times\mathbf m)\cdot\mathbf l],
\end{aligned}
\label{dc_ahe}
\end{equation}
where $\Delta\rho=\rho_{||}-\rho_{\perp}$ with $\rho_{||}$ ($\rho_{\perp}$)
being the longitudinal (transverse) resistivity of the HM/FM bilayer when
$\mathbf M$ is parallel (perpendicular) to $\mathbf J_\mathrm{FM}$, $R_1$
describes the AHE of the FM, and $\mathbf l$ is the displacement vector between
two electrode contact points used to measure the dc-voltage.

$U_\mathrm{ISHE}$ comes from the ISHE that converts a pure spin current $\mathbf J_s$
pumped by precessing magnetization near the st-FMR to a charge current.
The pumped spin current $\mathbf J_s$ is\cite{nakayama,bauer,tserkovnyak},
\begin{equation}
\mathbf J_s=\frac{\hbar}{4\pi}g_\mathrm{eff}^{\uparrow\downarrow}\frac{1}{M^2}
\mathbf M\times\frac{\partial\mathbf M}{\partial t},
\label{sp}
\end{equation}
where $g_\mathrm{eff}^{\uparrow\downarrow}$ is the effective spin mixing conductance.
$\mathbf J_s$ is then converted to an electric current in the HM layer,
\begin{equation}
\mathbf J_\mathrm{ISHE}=-\frac{2e}{\hbar}\theta_\mathrm{SH}\hat z\times\mathbf J_s
\label{j_ish}
\end{equation}
which results in a dc-voltage,
\begin{equation}
\begin{aligned}
U_\mathrm{ISHE} &= \frac{1}{\sigma_\mathrm{HM}}\langle\mathbf J_\mathrm{ISHE}\rangle\cdot\mathbf l\\
&= -\frac{2e}{\hbar}\frac{\theta_\mathrm{SH}}{\sigma_\mathrm{HM}}
(\hat z\times\langle\mathbf J_s\rangle)\cdot\mathbf l,
\end{aligned}
\label{dc_ish}
\end{equation}
where $\langle ... \rangle$ denotes the time average.
From Eq. \eqref{sp}, the dc spin current is
$\langle\mathbf J_s\rangle=g_\mathrm{eff}^{\uparrow\downarrow}\frac{\hbar\omega}{4\pi M^2}
\mathrm{Im}(m_X^\ast m_Y)\hat Z$. Thus, $U_\mathrm{ISHE}$ becomes
\begin{equation}
U_\mathrm{ISHE}=
-\frac{g_\mathrm{eff}^{\uparrow\downarrow}\theta_\mathrm{SH} e \omega}
{2\pi \sigma_\mathrm{HM} M^2}
\mathrm{Im}(m_X^\ast m_Y) [(z\times \hat Z)\cdot\mathbf l]
\label{dc_ish2}
\end{equation}
According to Eqs. \eqref{dc_amr}, \eqref{dc_ahe} and \eqref{dc_ish2},
the dc-voltage from the generalized Ohm's law and ISHE depend on how
$\mathbf m$ responds to $\mathbf h$ and $\mathbf e$, or the dynamic magnetic
susceptibility matrices $\tensor\chi$ and $\tensor\kappa$ near st-FMR.

\subsection{Dynamic magnetic susceptibility matrix $\tensor\chi$ and $\tensor\kappa$}

As mentioned in the last subsection, it is convenient to characterize the dynamical
component $\mathbf m$ by the dynamic susceptibilities $\tensor\chi$ and $\tensor\kappa$
as $\mathbf m=\tensor\chi\mathbf h+\tensor\kappa\mathbf e$, although $\mathbf j_
\mathrm{HM}$ generated by $\mathbf e$ is the ultimate source of $\mathbf h$ and rf SOT.
The universal form of $\tensor\chi(\omega)$ has been obtained in our previous work
\cite{wang2017}:
\begin{equation}
\tensor\chi(\omega)=
\frac{\pi\Gamma}{2}[L(\omega,\omega_0,\Gamma)+iD(\omega,\omega_0,\Gamma)]\tensor C,
\label{chi_f}
\end{equation}
where $L(\omega,\omega_0,\Gamma)$ is the Lorentzian function,
\begin{equation}
L(\omega, \omega_0,\Gamma)=
\frac{1}{\pi}\frac{\frac{\Gamma}{2}}{(\omega-\omega_0)^2+(\frac{\Gamma}{2})^2},
\label{lorentz}
\end{equation}
and the function $D(\omega, \omega_0,\Gamma)$ is
\begin{equation}
D(\omega,\omega_0,\Gamma)=
\frac{1}{\pi}\frac{\omega-\omega_0}{(\omega-\omega_0)^2+(\frac{\Gamma}{2})^2},
\label{disp}
\end{equation}
where $\omega_0$ denotes the resonance frequency and $\Gamma$ is the
linewidth which is a positive number. The matrix $\tensor C$ is
\begin{equation}
\tensor C=\left(
\begin{matrix}
iC_1 & C_3+iC_2 & 0\\
-C_3+iC_2 & iC_4 & 0\\
0 & 0 & 0\\
\end{matrix}
\right).
\label{C}
\end{equation}

In the case that the microwave frequency $\omega$ is fixed
and the applied static magnetic field $H$ is swept, the field-dependence
of $\tensor\chi$ has the following form for an arbitrary FM \cite{wang2017},
\begin{equation}
\tensor\chi(H)=\frac{\pi\Gamma_1}{2}[L(H,H_0,\Gamma_1)
+i\frac{-\zeta}{|\zeta|}D(H,H_0,\Gamma_1)]\tensor C(H_0),
\label{chi_H}
\end{equation}
where $H_0$ is the resonance field, $\zeta=\frac{d\omega_0}{dH}\big|_{H=H_0}$
and $\Gamma_1=\Gamma(H_0)/|\zeta|$ is the linewidth of the field.

Because $\mathbf e$ generates $\mathbf j_\mathrm{HM}$ that is the ultimate
source of $\mathbf h$ and rf SOT, $\tensor\kappa$ and $\tensor\chi$ are related.
To find the relationship between $\tensor\kappa$ and $\tensor\chi$, we start
from the generalized LLG equation \eqref{llgs_eq} which can be recasted as
\begin{equation}
\frac{\partial\mathbf M}{\partial t}=-\gamma\mathbf M\times(\mathbf H_\mathrm
{eff}+\mathbf H_\mathrm{st})+\frac{\alpha}{M}\mathbf M\times\frac{\partial\mathbf M}{\partial t},
\label{llgs_eq2}
\end{equation}
where $\mathbf H_\mathrm{st}$ denotes the effective field from the
SOT in Eq. \eqref{sot},
\begin{equation}
\mathbf H_\mathrm{st}=\frac{a}{M}\mathbf M\times\mathbf J_\mathrm s
-\beta a\mathbf J_\mathrm s.
\end{equation}
Because the spin current $\mathbf J_\mathrm s$ contains only one rf component,
up to the linear term in the precessing magnetization $\mathbf m$,
$\mathbf H_\mathrm{st}$ can be written as
\begin{equation}
\mathbf H_\mathrm{st}=\mathrm{Re}(\mathbf h_\mathrm{st}e^{-i\omega t}),
\end{equation}
where
\begin{equation}
\begin{aligned}
\mathbf h_\mathrm{st} =& -\frac{a}{M}\theta_\mathrm{SH}\sigma_\mathrm{HM}
(\frac{\hbar}{2e})\mathbf M_0\times(\hat z\times\mathbf e)\\
&+\beta a\theta_\mathrm{SH}\sigma_\mathrm{HM}(\frac{\hbar}{2e})\hat z\times\mathbf e.
\end{aligned}
\end{equation}
Thus, one can view st-FMR as the usual FMR under a total effective rf field of
$\mathbf h+\mathbf h_\mathrm{st}$, and the response of $\mathbf m$ is
\begin{equation}
\mathbf m=\tensor\chi(\mathbf h+\mathbf h_\mathrm{st})
=\tensor\chi\mathbf h+\tensor\kappa\mathbf e.
\end{equation}
$\tensor\kappa$ relates to $\tensor\chi$ as,
\begin{equation}
\tensor\kappa=\tensor\chi[-\frac{1}{M}\Lambda(\mathbf M_0)+\beta]
a\theta_\mathrm{SH}\sigma_\mathrm{HM}(\frac{\hbar}{2e})\Lambda(\hat z),
\label{kappa}
\end{equation}
where $\Lambda$ denotes the operator of cross product, i.e.,
$\Lambda(\mathbf a)\mathbf b=\mathbf a\times\mathbf b$.
Substituting the universal form of $\tensor\chi(\omega)$ into Eq. \eqref{kappa},
one can obtain the universal form of frequency-dependence of $\tensor\kappa$,
\begin{equation}
\tensor\kappa(\omega)=a\theta_\mathrm{SH}\sigma_\mathrm{HM}(\frac{\hbar}{2e})
\frac{\pi\Gamma}{2}[L(\omega,\omega_0,\Gamma)+iD(\omega,\omega_0,\Gamma)]\tensor C_e,
\label{kappa_f}
\end{equation}
where
\begin{widetext}
\begin{equation}
\tensor C_e=\left(
\begin{matrix}
[\beta C_3+i(C_1+\beta C_2)]\cos\theta & [C_3+i(C_2-\beta C_1)]\cos\theta
& [\beta C_3+i(C_1+\beta C_2)]\sin\theta\\
[-C_3+i(C_2+\beta C_4)]\cos\theta & [\beta C_3+i(C_4-\beta C_2)]\cos\theta
& [-C_3+i(C_2+\beta C_4)]\sin\theta\\
0 & 0 & 0\\
\end{matrix}
\right),
\label{C_e}
\end{equation}
\end{widetext}
with $\theta$ being the polar angle of $\mathbf M_0$ in the $xyz$-coordinate.

The field-dependence of $\tensor\kappa$ can be obtained by substituting
Eq. \eqref{chi_H} into Eq. \eqref{kappa},
\begin{equation}
\tensor\kappa(H) = a\theta_\mathrm{SH}\sigma_\mathrm{HM}(\frac{\hbar}{2e})
\frac{\pi\Gamma_1}{2}[L(H,H_0,\Gamma_1)
-i\frac{\zeta}{|\zeta|}D(H,H_0,\Gamma_1)]\tensor C_e.
\label{kappa_H}
\end{equation}

Consequently, the magnetization motion at st-FMR can be expressed as
\begin{widetext}
\begin{equation}
\left(
\begin{matrix}
m_X \\ m_Y \\ m_Z
\end{matrix}
\right)=\left(
\begin{matrix}
\chi_{XX} & \chi_{XY} & 0\\
\chi_{YX} & \chi_{YY} & 0\\
0 & 0 & 0
\end{matrix}
\right)\left(
\begin{matrix}
h_X \\ h_Y \\ h_Z
\end{matrix}
\right)
+\left(
\begin{matrix}
\kappa_{XX} & \kappa_{XY} & \kappa_{XZ}\\
\kappa_{YX} & \kappa_{YY} & \kappa_{YZ}\\
0 & 0 & 0
\end{matrix}
\right)\left(
\begin{matrix}
e_X \\ e_Y \\ e_Z
\end{matrix}
\right).
\label{chi}
\end{equation}
\end{widetext}
After the universal forms of $\tensor\chi$ and $\tensor\kappa$ are obtained,
one is able to find the dc-voltage signals attributed from the AMR, AHE and ISHE.

\subsection{The lineshape of dc-voltage}

Substituting Eq. \eqref{chi} into Eqs. \eqref{dc_amr}, \eqref{dc_ahe} and \eqref{dc_ish2},
$U_\mathrm{AMR}$, $U_\mathrm{AHE}$ and $U_\mathrm{ISHE}$ in terms of $\tensor\chi$
and $\tensor\kappa$ are
\begin{equation}
U_\mathrm{AMR}
=\frac{\Delta\rho}{2M}\mathrm{Re}[(j_{\mathrm{FM},i}^\ast l_Z+j_{\mathrm{FM},Z}^\ast l_i)
(\chi_{ij}h_j+\kappa_{ij}e_j)]
\label{dc_amr_chi}
\end{equation}
\begin{equation}
U_\mathrm{AHE}=
-\frac{R_1}{2}\mathrm{Re}[\epsilon_{ijk}j_{\mathrm{FM},j}^\ast l_i(\chi_{kl}h_l
+\kappa_{kl}e_l)],
\label{dc_ahe_chi}
\end{equation}
\begin{equation}
U_\mathrm{ISHE}=
-\frac{g_\mathrm{eff}^{\uparrow\downarrow}\theta_\mathrm{SH}e \omega l_Y \sin\theta}
{2\pi\sigma_\mathrm{HM}M^2}
\mathrm{Im}[(\chi_{Xi}^\ast h_i^\ast+\kappa_{Xi}^\ast e_i^\ast)
(\chi_{Yj} h_j+\kappa_{Yj} e_j)],
\label{dc_ish_chi}
\end{equation}
where subscript indices $i$, $j$, $k$ and $l$ can be $X$, $Y$ and $Z$.
$\epsilon_{ijk}$ is the Levi-Civita symbol, and the Einstein summation convention is used.
Whether a matrix element of $\tensor\chi$ or $\tensor\kappa$ is involved
in dc-voltage depends on the applied microwave fields and experimental setup.
Substituting Eqs. \eqref{chi_f}, \eqref{C}, \eqref{kappa_f}
and \eqref{C_e} into Eqs. \eqref{dc_amr_chi}-\eqref{dc_ish_chi},
the frequency-dependence of dc-voltage can be
expressed in terms of Lorentzian and $D$ functions,
\begin{equation}
U_\mathrm{AMR}(\omega)=A_1\frac{\pi\Gamma}{2}L(\omega,\omega_0,\Gamma)
+A_2\frac{\pi\Gamma}{2}D(\omega,\omega_0,\Gamma),
\label{dc_amr_f}
\end{equation}
\begin{equation}
U_\mathrm{AHE}(\omega)=A_3\frac{\pi\Gamma}{2}L(\omega,\omega_0,\Gamma)
+A_4\frac{\pi\Gamma}{2}D(\omega,\omega_0,\Gamma),
\label{dc_ahe_f}
\end{equation}
\begin{equation}
U_\mathrm{ISHE}(\omega)=A_5\frac{\pi\Gamma}{2}L(\omega,\omega_0,\Gamma).
\label{dc_ish_f}
\end{equation}
$A_1\sim A_5$ are
\begin{equation}
\begin{aligned}
A_1 =& \frac{\Delta\rho}{2M}\mathrm{Re}[(j_{\mathrm{FM},i}^\ast l_Z+j_{\mathrm{FM},Z}^\ast l_i)
(C_{ij} h_j \\ &+\frac{\hbar}{2e}a\theta_\mathrm{SH}C_{e,ij}j_{\mathrm{HM},j})],\\
A_2 =& -\frac{\Delta\rho}{2M}\mathrm{Im}[(j_{\mathrm{FM},i}^\ast l_Z+j_{\mathrm{FM},Z}^\ast l_i)
(C_{ij} h_j \\ &+\frac{\hbar}{2e}a\theta_\mathrm{SH}C_{e,ij}j_{\mathrm{HM},j})],\\
A_3 =& -\frac{R_1}{2}\mathrm{Re}[\epsilon_{ijk}j_{\mathrm{FM},j}^\ast l_i
(C_{kl} h_l + \frac{\hbar}{2e}a\theta_\mathrm{SH} C_{e,kl} j_{\mathrm{HM},l})],\\
A_4 =& \frac{R_1}{2}\mathrm{Im}[\epsilon_{ijk}j_{\mathrm{FM},j}^\ast l_i
(C_{kl} h_l + \frac{\hbar}{2e}a\theta_\mathrm{SH} C_{e,kl} j_{\mathrm{HM},l})],\\
A_5 =& -
\frac{g_\mathrm{eff}^{\uparrow\downarrow}\theta_\mathrm{SH} e\omega l_Y\sin\theta}
{2\pi \sigma_\mathrm{HM} M^2}
\mathrm{Im}[(C_{Xi}^\ast h_i^\ast+\frac{\hbar}{2e} a\theta_\mathrm{SH}
C_{e,Xi}^\ast j_{\mathrm{HM},i}^\ast)
\\ &\cdot(C_{Yj} h_j+\frac{\hbar}{2e} a\theta_\mathrm{SH} C_{e,Yj} j_{\mathrm{HM},j})],
\end{aligned}
\label{dc_amp}
\end{equation}
where subscript indices $i$, $j$, $k$ and $l$ are $x$, $y$ and $z$.
$C_{ij}$ (or $C_{e,ij}$) is the element of the $i$-th row and the
$j$-th column of matrix $\tensor C$ defined in Eq. \eqref{C} (or
matrix $\tensor C_e$ defined in Eq. \eqref{C_e}).
Starting from the universal forms of $\tensor\chi(H)$ and $\tensor\kappa(H)$,
one can also find the field-dependence of dc-voltage lineshapes,
\begin{equation}
U_\mathrm{AMR}(H)=A_1\frac{\pi\Gamma}{2}L(H,H_0,\Gamma_1)
-\frac{\zeta}{|\zeta|}A_2\frac{\pi\Gamma}{2}D(H,H_0,\Gamma_1),
\label{dc_amr_H}
\end{equation}
\begin{equation}
U_\mathrm{AHE}(H)=A_3\frac{\pi\Gamma}{2}L(H,H_0,\Gamma_1)
-\frac{\zeta}{|\zeta|}A_4\frac{\pi\Gamma}{2}D(H,H_0,\Gamma_1),
\label{dc_ahe_H}
\end{equation}
\begin{equation}
U_\mathrm{ISHE}(H)=A_5\frac{\pi\Gamma}{2}L(H,H_0,\Gamma_1).
\label{dc_ish_H}
\end{equation}
The results tell us that the general dc-voltage lineshape near the st-FMR
have a {\it symmetric component} of the Lorentzian function and an
{\it antisymmetric component} of the $D$ function.
$A_1\sim A_5$ are linear combinations of $C_1\sim C_4$ whose coefficients
depend on magnetic anisotropy and experimental setup, and their values
determine the relative weights of the symmetric and antisymmetric components.

\section{Results and discussion}

In this section, we use both easy-plane and biaxial models in two experimental
configurations to illustrate possible angle-dependence of dc-voltage and dc-voltage
lineshape. We will also propose a proper way to experimentally determine
spin-Hall angle of the HM.

\begin{figure}
\centering
\includegraphics[width=0.4\textwidth]{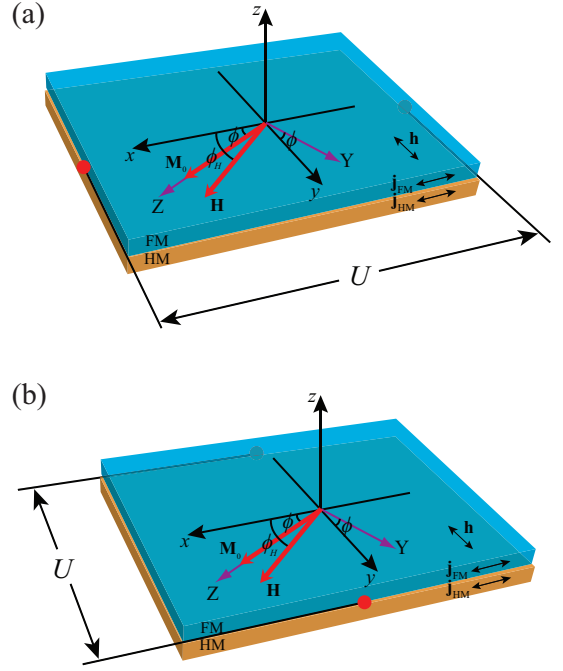}\\
\caption{Schematic illustration for two experimental configurations of st-FMR.
A HM/FM bilayer lies in the $xy$-plane. The stable magnetization $\mathbf M_0$
is in the sample plane by applying an in-plane static magnetic field $\mathbf H$.
$\phi$ is the angle between $\mathbf M_0$ and the $x$-axis, and $\phi_H$ is the
angle between $\mathbf H$ and the $x$-axis. The definitions of the $xyz$- and
$XYZ$-coordinates are the same as Fig. \ref{configs_chi} with $\theta=90^\mathrm o$.
The rf electric currents $\mathbf j_\mathrm{FM}$ and $\mathbf j_\mathrm{HM}$
are along the $x$-direction, and the rf Oersted field $\mathbf h$ is
along the $y$-direction according to the Ampere's law. The displacement
vector between two electrodes is along the $x$-direction (a) or the
$y$-direction (b).
}
\label{configs_angle}
\end{figure}

Our model system, which mimics popular experimental setups, is shown in Fig.
\ref{configs_angle}. A HM/FM bilayer film lies in the $xy$-plane with the
length $l_x$ along the $x$-direction and width $l_y$ along the $y$-direction.
The rf current $\mathbf j_a$ ($a=$FM, HM) is along the $x$-axis.
The effective magnetic field is
\begin{equation}
\mathbf H_\mathrm{eff}=\mathbf H+K_xM_x\hat x-M_z\hat z
+\mathrm{Re}(\mathbf h e^{-i\omega t}),
\label{h_eff}
\end{equation}
where the static in-plane magnetic field $\mathbf H$ is with a $\phi_H$ angle about
the $x$-axis, the second and third terms are respectively the easy-axis and hard-axis
(shape) anisotropy fields, and the forth term is the rf Oersted field.
In the following analyses of two experimental configurations, we firstly consider
easy-plane case of $K_x=0$, and then expand the results to the biaxial case of $K_x>0$.

\subsection{Dc-voltage along the rf current}

In this configuration as shown in Fig. \ref{configs_angle}(a), the dc-voltage
is measured along the direction of rf electric current, i.e. $\mathbf l = l_x\hat x$.
The dc-voltage near the FMR comes from the AMR and ISHE because the dc electric
field from AHE is transverse to the rf current. According to the spin pumping and
ISHE, the dc current $\langle \mathbf J_\mathrm{ISHE}\rangle$ near the FMR is in the
sample plane and orthogonal to $\mathbf M_0$. It would be zero when $\mathbf M_0$ is
along $\hat y$. Thus, $\langle \mathbf J_\mathrm{ISHE}\rangle$ has a $x$-component and
results in a dc-voltage along $\hat x$ only when $\mathbf M_0$ deviates from the $x$-
and $y$-directions.

\subsubsection{Easy-plane case}

For an easy-plane FM film where the $z$-axis is the hard axis of
the film, the stable magnetization $\mathbf M_0$ in the absence of
microwave field is collinear with $\mathbf H$, i.e. $\phi=\phi_H$.
According to Eq. \eqref{llgs_eq}, the linearized LLG equation
in the present case becomes
\begin{equation}
\begin{gathered}
-i\omega\mathbf m=-\gamma\mathbf m\times\mathbf H
-\gamma\mathbf M_0\times(\mathbf h -m_{z}\hat z)\\
-i\omega\frac{\alpha}{M}\mathbf M_0\times\mathbf m
+\gamma\frac{a}{M}(\frac{\hbar}{2e})\theta_\mathrm{SH}\sigma_\mathrm{HM}
\mathbf M_0\times[\mathbf M_0\times(\hat z\times \mathbf e)],
\end{gathered}
\label{linear_llgs_eq}
\end{equation}
where we assume that the field-like torque is very small and can be
neglected, i.e. $\beta=0$.
The exact solution of this equation allows us to obtain the expressions
of $H_0$, $\Gamma_1$, $C_1$, $C_2$, $C_3$ and $C_4$ which determine
$\tensor\chi$ and $\tensor\kappa$ for an easy-plane model.

In the absence of any driving force and damping, from Eq.
\eqref{linear_llgs_eq} it is easy to find the FMR frequency
$\omega_0=\gamma\sqrt{H(M+H)}$ which is the well-known Kittel's formula.
Thus, the resonance field $H_0$ for a given microwave frequency $\omega$
can be obtained as
\begin{equation}
H_0=\sqrt{\frac{\omega^2}{\gamma^2}+\frac{M^2}{4}}-\frac{M}{2}.
\label{kittel}
\end{equation}
To find the linewidth $\Gamma_1$ and the real numbers $C_1\sim C_4$,
we start from the non-zero matrix elements of $\tensor\chi$:
\begin{equation}
\begin{aligned}
\chi_{XX} &= \frac{\gamma M(-\gamma H+i\alpha\omega)}{\omega^2-\omega_0^2
+i\alpha\gamma\omega(M+2H)},\\
\chi_{XY} &= -\chi_{YX} = \frac{i\omega\gamma M}{\omega^2-\omega_0^2
+i\alpha\gamma\omega(M_s+2H)},\\
\chi_{YY} &= \frac{\gamma M(-\gamma H -\gamma M+i\alpha\omega)}
{\omega^2-\omega_0^2 +i\alpha\gamma\omega(M+2H)}.
\end{aligned}
\label{chi_uni}
\end{equation}
Eq. \eqref{chi_uni} can be written as the sum of a Lorentzian function
and an D function near the resonance field $H_0$. In terms of parameters
defined in Eq. \eqref{kappa_f}-\eqref{kappa_H}, it is easy to obtain
\begin{equation}
\Gamma_1=\frac{2\alpha\omega}{\gamma},
\label{width_uni}
\end{equation}
and
\begin{equation}
\begin{aligned}
C_1 &= \frac{\gamma M H_0}{\alpha\omega(2H_0+M)},\\
C_2 &= 0, \\
C_3 &= \frac{M}{\alpha(2H_0+M)},\\
C_4 &= \frac{\gamma M (H_0+M)}{\alpha\omega(2H_0+M)},
\end{aligned}
\label{C_uni}
\end{equation}
where Eq. \eqref{width_uni} is usually used in experiments to determine
the Gilbert damping coefficient $\alpha$. From the Kittel's formula, it is
obvious that $\zeta>0$ which results in a minus sign in front of $D$
function in Eqs. \eqref{chi_H}, \eqref{dc_amr_H} and \eqref{dc_ahe_H}.
It is only true for an easy-plane model in which $H_0$, $\Gamma_1$ and
$C_1\sim C_4$ does not depend on $\phi_H$ for an in-plane field $\mathbf H$.
For a biaxial model where $K_x>0$, all these parameters depend on
$\phi_H$ in general.

In the $XYZ$-coordinate, the displacement $\mathbf l=l_x\hat x$ between
the two electrodes becomes
\begin{equation}
\mathbf l=-l_x\sin\phi\hat Y+l_x\cos\phi\hat Z.
\label{lx}
\end{equation}
The dc-voltage from each contribution can be obtained by substituting
Eqs. \eqref{kittel}-\eqref{C_uni} and Eq. \eqref{lx} into Eq. \eqref{dc_amp},
\begin{equation}
\begin{aligned}
A_1 =& \frac{\Delta\rho j_\mathrm{HM}j_\mathrm{FM}l_x}
{2\alpha(2H_0+M)}\cdot a\theta_\mathrm{SH}\frac{\hbar}{2e}\cdot\sin2\phi_H\cos\phi_H, \\
A_2 =& \frac{\Delta\rho j_\mathrm{HM}j_\mathrm{FM}l_x\omega}
{2\alpha\gamma H_0(2H_0+M)}\cdot \frac{d_\mathrm{HM}}{2}\cdot\sin2\phi_H\cos\phi_H, \\
A_3 =& A_4 = 0, \\
A_5 =& \frac{g_\mathrm{eff}^{\uparrow\downarrow}\theta_\mathrm{SH}ej_\mathrm{HM}^2l_x \gamma H_0 }
{4\pi\alpha^2\sigma_\mathrm{HM}(2H_0+M)^2}\cdot
[(a\theta_\mathrm{SH}\frac{\hbar}{2e})^2+(\frac{\omega d_\mathrm{HM}}{2\gamma H_0})^2]\\
&\cdot\sin2\phi_H\cos\phi_H. \\
\end{aligned}
\label{dc_amp_lx}
\end{equation}
$A_1$ and $A_2$ are respectively from the AMR contribution due to the rf SOT driven
and rf Oersted field driven magnetization motion. Thus, $A_1$ is proportional to
$\mathbf j_s$ converted from $\mathbf j_\mathrm{HM}$ via SHE that, in turn, is
proportional to $\theta_\mathrm{SH}$, while $A_2$ is independent of $\theta_\mathrm{SH}$.
Both $A_3$ and $A_4$ are zero due to the absence of the AHE contribution as mentioned before.
The two terms in $A_5$ depend on $\theta_\mathrm{SH}$. One is linear in $\theta_\mathrm{SH}$
because of ISHE. The other is proportional to its cubic form.
This is because the spin current is proportional to the square of amplitude of magnetization
deviation that, in turn, come from both rf Oersted field that does not depends on
$\theta_\mathrm{SH}$ and the effective field generated by
rf SOT that is proportional to $\theta_\mathrm{SH}$ due to SHE.

Equation \eqref{dc_amp_lx} indicates that both symmetric and antisymmetric
components of dc-voltage lineshapes follow the same angle-dependence of
$\sin2\phi_H\cos\phi_H$. In the previous estimation of the spin-Hall angle $\theta
_\mathrm{SH}$, $U_\mathrm{ISHE}$ is assumed to be negligible, i.e. $A_5=0$, for the
reason that $U_\mathrm{ISHE}$ is high order in the spin-Hall angle \cite{buhrman}.
Thus, the symmetric component of dc-voltage signal is completely from the AMR.
Under this assumption, the spin-Hall angle can be estimated by $\theta_\mathrm{SH}
=\frac{S}{A}\cdot\frac{\omega d_\mathrm{HM}e}{a\gamma H_0\hbar}$
where $S$ and $A$ are respectively the amplitudes of symmetric
and antisymmetric components of dc-voltage lineshape for any angle $\phi_H$.
However, the estimated value by this approach is found to be overestimated
compared with spin pumping experiments \cite{hoffmann,azevedo,hfding,ohno},
which indicates that this assumption is questionable.

A more precise estimation of the spin-Hall angle $\theta_\mathrm{SH}$
can be obtained by taking into account the ISHE contribution of dc-voltage.
According to Eq. \eqref{dc_amp_lx}, $U_\mathrm{ISHE}$ has the exact same
symmetry and angle-dependence as the AMR contribution to dc-voltage due to
SOT, however, it will not prevent one from obtaining $\theta_\mathrm{SH}$.
Starting from Eq. \eqref{dc_amp_lx}, the ratio $S/A$, where $S=A_1+A_3+A_5$
and $A=A_2+A_4$, is
\begin{equation}
\frac{S}{A} = a_1 \theta_\mathrm{SH} + a_2 \theta_\mathrm{SH}^3,
\label{ratio_lx}
\end{equation}
where $S/A$ is measured in experiments, and $a_1$ and $a_2$ are
\begin{equation}
\begin{aligned}
a_1 &= \frac{a\gamma H_0\hbar}{\omega d_\mathrm{HM} e}
+\frac{g_\mathrm{eff}^{\uparrow\downarrow} e \omega d_\mathrm{HM}}
{4\pi\alpha(2H_0+M)\Delta\rho\sigma_\mathrm{FM}}\\
a_2 &=
\frac{a^2 g_\mathrm{eff}^{\uparrow\downarrow}\gamma^2H_0^2\hbar^2}
{4\pi\alpha(2H_0+M)\Delta\rho\sigma_\mathrm{FM}\omega d_\mathrm{HM}e}.
\end{aligned}
\end{equation}
Consequently, the corrected value of $\theta_\mathrm{SH}$ can be determined
from Eq. \eqref{ratio_lx} since $\theta_\mathrm{SH}$ is the only unknown.
Different from the previous argument \cite{buhrman}, two terms on the right-hand side
of Eq. \eqref{ratio_lx} are in general of the same order for typical materials so
that $\theta_\mathrm{SH}$ is not proportional to the ratio $S/A$ as claimed before.

\begin{figure}
\centering
\includegraphics[width=0.39\textwidth]{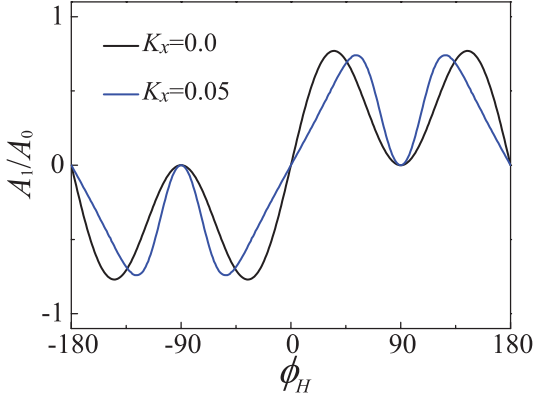}\\
\caption{Angle-dependence of $A_1$ in units of $A_0=\frac{\Delta\rho j_\mathrm{HM}
j_\mathrm{FM}l_x a\theta_\mathrm{SH}}{2\alpha(2H_0+M)}\cdot \frac{\hbar}{2e}$
for the setup shown in Fig. \ref{configs_angle}(a).
The model parameters are $\omega=9.0$ GHz, $M=8.0\times 10^5$ A/m and easy-axis
anisotropy coefficient $K_x=0.0$ (black curve) or $K_x=0.05$ (blue curve).
The black curve is plotted according to Eq. \eqref{dc_amp_lx}, and the
blue curve is numerically calculated from Eq. \eqref{dc_amp_lx2}.
}
\label{angle_lx}
\end{figure}

\subsubsection{Biaxial case}

For a general biaxial case with the easy-axis anisotropy coefficient $K_x>0$, the
static magnetization $\mathbf M_0$ in the absence of microwave fields is non-collinear
to static magnetic field $\mathbf H$, i.e. $\phi\neq\phi_H$ but $\phi=\phi(\phi_H)$.
$H_0(\phi_H)$, $\Gamma_1(\phi_H)$, $C_1(\phi_H)$, $C_3(\phi_H)$ and $C_4(\phi_H)$
are all functions of $\phi_H$, which can be numerically obtained once the material
parameters are given or be determined by standard microwave absorption measurements
\cite{wang2017}. Notice that $C_2=0$ is still satisfied because the energy density
function corresponding to the effective field of Eq. \eqref{h_eff} is symmetric
about the $YZ$-plane (or the $xy$-plane) \cite{wang2017}. Consequently, from Eq.
\eqref{dc_amp} one can obtain the dc-voltage for a biaxial model as follows,
\begin{equation}
\begin{aligned}
A_1 =& a_\mathrm{AMR}\cdot C_3(\phi_H)\sin2\phi(\phi_H)\cos\phi(\phi_H),\\
A_2 =& b_\mathrm{AMR}\cdot C_4(\phi_H)\sin2\phi(\phi_H)\cos\phi(\phi_H),\\
A_3 =& A_4 = 0, \\
A_5 =& [a_\mathrm{ISHE,1}\cdot C_1(\phi_H)+a_\mathrm{ISHE,2}\cdot C_4(\phi_H)] \\
& \cdot C_3(\phi_H)\sin2\phi(\phi_H)\cos\phi(\phi_H),
\end{aligned}
\label{dc_amp_lx2}
\end{equation}
where
\begin{equation}
\begin{aligned}
a_\mathrm{AMR} =& \frac{\Delta\rho j_\mathrm{HM}j_\mathrm{FM}l_x}
{2M}\cdot a\theta_\mathrm{SH}\frac{\hbar}{2e},\\
b_\mathrm{AMR} =& \frac{\Delta\rho j_\mathrm{HM}j_\mathrm{FM}l_x}
{2M}\cdot\frac{d_\mathrm{HM}}{2},\\
a_\mathrm{ISHE,1} =&
\frac{g_\mathrm{eff}^{\uparrow\downarrow}\theta_\mathrm{SH} e\omega j_\mathrm{HM}^2l_x}
{4\pi \sigma_\mathrm{HM} M^2}\cdot(a\theta_\mathrm{SH}\frac{\hbar}{2e})^2, \\
a_\mathrm{ISHE,2} =&
\frac{g_\mathrm{eff}^{\uparrow\downarrow}\theta_\mathrm{SH} e\omega j_\mathrm{HM}^2l_x}
{4\pi \sigma_\mathrm{HM} M^2}\cdot(\frac{d_\mathrm{HM}}{2})^2.
\end{aligned}
\label{dc_amp2_lx2}
\end{equation}
Obviously, the angle-dependence of dc-voltage in a biaxial model no longer follows
$\sin2\phi_H\cos\phi_H$. The angle-dependences of different components are different
in a biaxial model because $A_1$ and $A_2$ are respectively proportional to
$C_3(\phi_H)$ and $C_4(\phi_H)$, and $A_5$ is proportional to linear combinations of
$C_1(\phi_H)C_3(\phi_H)$ and $C_3(\phi_H)C_4(\phi_H)$.
This allows one to separate various contributions to dc-voltage. Figure \ref{angle_lx}
shows the angle-dependence of $A_1$ for an easy-plane ($K_x=0$) model (black curve) and a
biaxial model of $K_x=0.05$ (blue curve) for $\omega=9.0$ GHz and $M=8.0\times 10^5$ A/m.
The black and blue curves are respective plots of Eq. \eqref{dc_amp_lx} and  Eq. \eqref
{dc_amp_lx2}. For both cases, $A_1$ is in the units of $A_0=\frac{\Delta\rho j_\mathrm{HM}
j_\mathrm{FM}l_x a\theta_\mathrm{SH}}{2\alpha(2H_0+M)}\cdot \frac{\hbar}{2e}$. The angle-dependence
of $A_1$ for an easy-plane model follows $\sin2\phi_H\cos\phi_H$, while, for a
biaxial model of $K_x=0.05$, it apparently deviates from $\sin2\phi_H\cos\phi_H$.
Thus, one can tell whether or not an
FM is a biaxial magnetic film by looking at the angle-dependence of dc-voltage.

From Eqs. \eqref{dc_amp_lx2} and \eqref{dc_amp2_lx2}, following recipe can be used to
determine the spin-Hall angle.

\emph{Step I} Determine the angle-dependence of $C_i$ ($i=1,2,3,4$)
by standard microwave absorption experiments \cite{wang2017}.

\emph{Step II} After $C_i(\phi_H)$ ($i=1,2,3,4$) is obtained,
determine $a_\mathrm{AMR}$, $b_\mathrm{AMR}$, $a_\mathrm{ISHE,1}$
and $a_\mathrm{ISHE,2}$ by fitting the experimental curves
according to Eq. \eqref{dc_amp_lx2}.

\emph{Step III} The spin-Hall angle $\theta_\mathrm{SH}$ can be determined by
\begin{equation}
\theta_\mathrm{SH}=\frac{d_\mathrm{HM}}{a}\frac{e}{\hbar}
\frac{a_\mathrm{AMR}}{b_\mathrm{AMR}}.
\end{equation}

From the above steps, the spin-Hall angle can be determined for a biaxial sample
in the experimental setup of Fig. \ref{configs_angle}(a).

\subsection{Dc-voltage transverse to the rf current}

Fig. \ref{configs_angle}(b) is another widely used experimental configuration in
which the dc-voltage is measured transverse to rf current direction.
Different from the configuration in Fig. \ref{configs_angle}(a),
AMR, AHE and ISHE will all contribute to dc-voltage in this case.
Since the AHE generates a dc electric field transverse to rf current, it
should be very important in the present configuration. As mentioned before that
$\langle \mathbf J_\mathrm{ISHE}\rangle$ is in the sample plane and orthogonal to
$\mathbf M_0$, $\langle \mathbf J_\mathrm{ISHE}\rangle$ has in general a
$y$-component and can result in a dc-voltage along $\hat y$ when $\mathbf M_0$
is not parallel to $\hat y$.

\subsubsection{Easy-plane case}

For an easy-plane model, $H_0$, $\Gamma_1$, and $C_1\sim C_4$ are given by  Eqs.
\eqref{kittel}-\eqref{C_uni}.
In the $XYZ$-coordinate, the displacement $\mathbf l=l_y\hat y$ is
\begin{equation}
\mathbf l=l_y\cos\phi\hat Y+l_y\sin\phi\hat Z.
\label{ly}
\end{equation}
Substituting Eqs. \eqref{kittel}-\eqref{C_uni} and Eq. \eqref{ly}
into Eq. \eqref{dc_amp}, the amplitude of each dc-voltage component is,
\begin{equation}
\begin{aligned}
A_1 =& a_\mathrm{AMR}\cdot\cos2\phi_H\cos\phi_H, \\
A_2 =& b_\mathrm{AMR}\cdot\cos2\phi_H\cos\phi_H, \\
A_3 =& a_\mathrm{AHE}\cdot\cos\phi_H, \\
A_4 =& b_\mathrm{AHE}\cdot\cos\phi_H, \\
A_5 =& a_\mathrm{ISHE}\cdot\cos^3\phi_H,
\end{aligned}
\label{dc_amp_ly}
\end{equation}
where
\begin{equation}
\begin{aligned}
a_\mathrm{AMR} &= -\frac{\Delta\rho j_\mathrm{HM}j_\mathrm{FM}l_y}
{2\alpha(2H_0+M)}
\cdot a\theta_\mathrm{SH}\frac{\hbar}{2e},\\
b_\mathrm{AMR} &= -\frac{\Delta\rho j_\mathrm{HM}j_\mathrm{FM}l_y\omega}
{2\alpha\gamma H_0(2H_0+M)}
\cdot\frac{d_\mathrm{HM}}{2},\\
a_\mathrm{AHE} &= \frac{R_1j_\mathrm{HM}j_\mathrm{FM}l_y M}
{2\alpha(2H_0+M)}
\cdot\frac{d_\mathrm{HM}}{2},\\
b_\mathrm{AHE} &= -\frac{R_1j_\mathrm{HM}j_\mathrm{FM}l_y\gamma MH_0}
{2\alpha\omega(2H_0+M)}
\cdot a\theta_\mathrm{SH}\frac{\hbar}{2e},\\
a_\mathrm{ISHE} &= -
\frac{g_\mathrm{eff}^{\uparrow\downarrow}\theta_\mathrm{SH} e j_\mathrm{HM}^2l_y\gamma H_0}
{2\pi\alpha^2\sigma_\mathrm{HM}(2H_0+M)^2}
[(a\theta_\mathrm{SH}\frac{\hbar}{2e})^2+(\frac{\omega d_\mathrm{HM}}{2\gamma H_0})^2].
\end{aligned}
\label{dc_amp2_ly}
\end{equation}
$a_\mathrm{AMR}$ and $b_\mathrm{AHE}$ are respectively from the AMR
and AHE due to the rf SOT driven magnetization motion. Thus, they are
proportional to $\mathbf j_s$ converted from $\mathbf j_\mathrm{HM}$
via SHE and is proportional to $\theta_\mathrm{SH}$. On the other hand,
$b_\mathrm{AMR}$ and $a_\mathrm{AHE}$ are respectively from the AMR and AHE
due to the rf Oersted field driven magnetization motion, so they are related
to neither the SHE nor ISHE and are independent of $\theta_\mathrm{SH}$.
$a_\mathrm{ISHE}$ has two terms which are respectively proportional
to $\theta_\mathrm{SH}^3$ and $\theta_\mathrm{SH}$ for the similar reason
mentioned below Eq. \eqref{dc_amp_lx}.

Different from the previous case, the angle-dependence of dc-voltages from the AMR,
AHE and ISHE are not the same in the present configuration.
The issue is then how to determine $a_\mathrm{AMR}$, $b_\mathrm{AMR}$,
$a_\mathrm{AHE}$, $b_\mathrm{AHE}$ and $a_\mathrm{ISHE}$ to distinguish
each contribution to dc-voltage and find the spin-Hall angle $\theta_\mathrm{SH}$.
The symmetric component contains three different angle-dependences:
$\cos2\phi\cos\phi$, $\cos\phi$ and $\cos^3\phi$, however, these three
functions are not linearly independent. Thus, a symmetric curve
cannot uniquely determine the coefficients, and we should start from
the antisymmetric part where the angle-dependences $\cos2\phi\cos\phi$
and $\cos\phi$ are linearly independent with each other.

From Eqs. \eqref{dc_amp_ly}-\eqref{dc_amp2_ly}, following recipe can be
used to distinguish each dc-voltage contribution and determine the spin-Hall
angle.

\emph{Step I} Fit the angle-dependence of antisymmetric component of dc-voltage
by $\cos2\phi\cos\phi$ and $\cos\phi$. The fitting numbers of $\cos2\phi\cos\phi$
and $\cos\phi$ are $b_\mathrm{AMR}$ and $b_\mathrm{AHE}$, respectively.

\emph{Step II} $a_\mathrm{AMR}$ and $a_\mathrm{AHE}$ can be determined
by $a_\mathrm{AMR}=\frac{\Delta\rho}{R_1}\frac{\omega}{\gamma MH_0}b_\mathrm{AHE}$
and $a_\mathrm{AHE}=-\frac{R_1}{\Delta\rho}\frac{\gamma MH_0}{\omega}b_\mathrm{AMR}$.

\emph{Step III} Subtracting $a_\mathrm{AMR}$- and $a_\mathrm{AHE}$-terms
from the symmetric component of dc-voltage, the rest part comes from
ISHE and can determine $a_\mathrm{ISHE}$ using Eq. \eqref{dc_amp_ly}.

\emph{Step IV} The spin-Hall angle can be determined by,
\begin{equation}
\theta_\mathrm{SH}=\frac{\omega d_\mathrm{HM}e}{a\gamma H_0\hbar}
\frac{a_\mathrm{AMR}}{b_\mathrm{AMR}}
=-\frac{\omega d_\mathrm{HM}e}{a\gamma H_0\hbar}
\frac{b_\mathrm{AHE}}{a_\mathrm{AHE}}.
\label{theta_sh_y}
\end{equation}

From the above steps, the dc-voltage from each source can be distinguished
and the spin-Hall angle can be determined for an easy-plane sample
in the setup of Fig. \ref{configs_angle}(b).

According to Eq. \eqref{dc_amp2_ly}, one has
\begin{equation}
\frac{a_\mathrm{AMR}}{b_\mathrm{AMR}}
=-\frac{b_\mathrm{AHE}}{a_\mathrm{AHE}}.
\label{ratio}
\end{equation}
This measures the ratio of the rf SOT and the torque by rf Oersted field.
If Eq. \eqref{ratio} cannot be satisfied after one extracts all numbers from an
experiment, it may indicate additional effects beyond the current model, e.g.
the extraordinary galvanomagnetic effects in polycrystalline magnetic films
\cite{wang2016}, or the longitudinal ISHE which will be discussed in the next
section.

\subsubsection{Biaxial case}

We consider a biaxial model in the configuration of Fig. \ref{configs_angle}(b).
The model is the same as the biaxial model used in the configuration of Fig.
\ref{configs_angle}(a) but replace $\mathbf l=l_x\hat x$ by $\mathbf l=l_y\hat y$.
By applying the biaxial model into the universal forms of dynamic magnetic
susceptibility and dc-voltage lineshape, one can obtain,
\begin{equation}
\begin{aligned}
A_1 =& a_\mathrm{AMR}\cdot C_3(\phi_H)\cos2\phi(\phi_H)\cos\phi(\phi_H), \\
A_2 =& b_\mathrm{AMR}\cdot C_4(\phi_H)\cos2\phi(\phi_H)\cos\phi(\phi_H), \\
A_3 =& a_\mathrm{AHE}\cdot C_3(\phi_H)\cos\phi(\phi_H), \\
A_4 =& b_\mathrm{AHE}\cdot C_1(\phi_H)\cos\phi(\phi_H), \\
A_5 =& [a_\mathrm{ISHE,1}\cdot C_1(\phi_H)+a_\mathrm{ISHE,2}\cdot C_4(\phi_H)] \\
& \cdot C_3(\phi_H)\cos^3\phi(\phi_H),
\end{aligned}
\label{dc_amp_ly2}
\end{equation}
where
\begin{equation}
\begin{aligned}
a_\mathrm{AMR} &= -\frac{\Delta\rho j_\mathrm{HM}j_\mathrm{FM}l_y}
{2M}\cdot a\theta_\mathrm{SH}\frac{\hbar}{2e},\\
b_\mathrm{AMR} &= -\frac{\Delta\rho j_\mathrm{HM}j_\mathrm{FM}l_y}
{2M}\cdot\frac{d_\mathrm{HM}}{2},\\
a_\mathrm{AHE} &= \frac{R_1j_\mathrm{HM}j_\mathrm{FM}l_y}
{2}\cdot\frac{d_\mathrm{HM}}{2},\\
b_\mathrm{AHE} &= -\frac{R_1j_\mathrm{HM}j_\mathrm{FM}l_y}
{2}\cdot a\theta_\mathrm{SH}\frac{\hbar}{2e},\\
a_\mathrm{ISHE,1} &=
-\frac{g_\mathrm{eff}^{\uparrow\downarrow}\theta_\mathrm{SH} e\omega j_\mathrm{HM}^2l_y}
{2\pi \sigma_\mathrm{HM} M^2}\cdot (a\theta_\mathrm{SH}\frac{\hbar}{2e})^2, \\
a_\mathrm{ISHE,2} &=
-\frac{g_\mathrm{eff}^{\uparrow\downarrow}\theta_\mathrm{SH} e\omega j_\mathrm{HM}^2l_y}
{2\pi \sigma_\mathrm{HM} M^2}\cdot(\frac{d_\mathrm{HM}}{2})^2,
\end{aligned}
\label{dc_amp2_ly2}
\end{equation}
from which it is obvious that the angle-dependence differs from that of the
easy-plane model. For the similar reasons mentioned after Eq. \eqref{dc_amp2_ly},
$a_\mathrm{AMR}$ and $b_\mathrm{AHE}$ are proportional to $\theta_\mathrm{SH}$
while $b_\mathrm{AMR}$ and $a_\mathrm{AHE}$ are independent of $\theta_\mathrm{SH}$.
Two terms from the ISHE are characterized by $a_\mathrm{ISHE,1}$ and
$a_\mathrm{ISHE,2}$.

From Eqs. \eqref{dc_amp_ly2}-\eqref{dc_amp2_ly2}, following recipe can be
used to separate dc-voltage signals from different contributions and to extract the
spin-Hall angle.

\emph{Step I} Determine the angle-dependence of $C_i$ ($i=1,2,3,4$)
by standard microwave absorption experiments \cite{wang2017}.

\emph{Step II} After $C_i(\phi_H)$ ($i=1,2,3,4$) is obtained,
determine $a_\mathrm{AMR}$, $b_\mathrm{AMR}$, $a_\mathrm{AHE}$,
$b_\mathrm{AHE}$, $a_\mathrm{ISHE,1}$
and $a_\mathrm{ISHE,2}$ by fitting the experimental curves
according to Eq. \eqref{dc_amp_ly2}.

\emph{Step III} The spin-Hall angle $\theta_\mathrm{SH}$ can be still determined
by Eq. \eqref{theta_sh_y}.

From the above steps, the dc-voltage from each source can be distinguished
and the spin-Hall angle can be determined for a biaxial
sample in the experimental configuration of Fig. \ref{configs_angle}(b).

Again, one can use Eq. \eqref{ratio} to test the model. If the extract model parameters
do not satisfy Eq. \eqref{ratio}, then there may exist other sources for the dc-voltage
like the extraordinary galvanomagnetic effects \cite{wang2016}, or the longitudinal
ISHE discussed below.

\subsection{Discussion}

So far, the electric current density converted from a spin current of polarization
$\vec p$ and magnitude $J^\mathrm{s}$ flowing along the $z$-direction via the
ISHE is assumed to be $\theta_\mathrm{SH}J^\mathrm{s}\hat z\times\vec{p}$.
In magnetic materials, however, the general physics principle can allow other types
of electric current density when the spin current interacts with the magnetization.
In the linear responses to $J^\mathrm{s}$, one can also construct the other charge
current density vector out of spin current $\tensor J^{\mathrm s}$ of a tensor of
rank 2 and magnetization $\mathbf M$. 
Let us denote $J_{ij}^\mathrm{s}$ as the spin current of polarization along the $j$-direction
($\vec p$) and flowing along the $i$-direction ($\hat z$ in the current case).

Similar to the derivation of AMR \cite{wang2016} under the assumption of physics
law being coordinate independent, the most general charge current density
$\mathbf J^{c}$ of a vector converted from a spin current $\tensor J^{\mathrm s}$,
within the linear response, should be
\begin{equation}
J_k^c =\frac{2e}{\hbar}\theta_{ijk}^\mathrm{SH} J_{ij}^\mathrm{s},
\nonumber
\end{equation}
where $J_k^{c}$ is the charge current density along the $k$-direction and
$\theta_{ijk}^\mathrm{SH}$ is the $ijk$-component of the general spin Hall angle
tensor $\tensor{\theta}^\mathrm{SH}$ of rank 3 that depends on $\mathbf M$.
$i,j,k=1,2,3$ stands for $x$-, $y$-, and $z$-directions and the Einstein
summation convention is assumed.
The most general form of $\theta_{ijk}^\mathrm{SH}$ is
\begin{equation}
\begin{aligned}
\theta_{ijk}^\mathrm{SH}=&\theta_0^\mathrm{SH}\epsilon_{ijk}
+\theta_1^\mathrm{SH}M_l\epsilon_{iln}\epsilon_{jnk}
+\theta_2^\mathrm{SH}M_lM_n\epsilon_{ilp}
\epsilon_{jpq}\epsilon_{kqn}\\
&+\theta_3^\mathrm{SH}M_iM_jM_k,
\label{gen-sh}
\end{aligned}
\end{equation}
where $\epsilon_{ijk}$ is the usual Levi-Civita symbol.
$\theta_0^\mathrm{SH}=\theta_\mathrm{SH}$ is the usual spin Hall angle that
does not interact with $\mathbf M$, $\theta_\alpha^\mathrm{SH}$ ($\alpha=1,2,3$)
that are respectively linear, quadratic and cubic in $\mathbf M$.
In the following, we limit ourselves to the first two terms,
and discuss the generated charge current
in two cases: (1) the spin current flows along its polarization $\vec p$,
and (2) the spin current flows transverse to its polarization $\vec p$.

\begin{figure}
\centering
\includegraphics[width=0.5\textwidth]{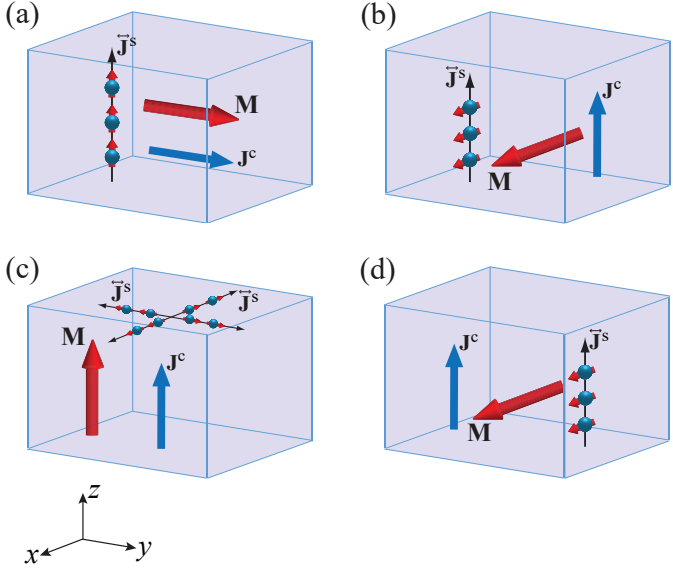}\\
\caption{Schematics of new ISHE ((a) and (b)) and SHE ((c) and (d))
in magnetic materials. (a) For a spin current flowing along the spin polarization
direction, a charge current can be generated along the magnetization perpendicular
to spin flowing direction (as well as the spin polarization).
(b) For a spin current flowing perpendicular to the spin polarization direction,
a charge current flows along the spin flowing direction when the magnetization is
along the spin polarization direction.
(c) For a charge current flowing along the magnetization, the generated spin current
flows perpendicular to the charge current and the spin polarization is along
the spin flow direction.
(d) For a charge current flowing perpendicular to the magnetization, the generated
spin current is along the charge current and the spin polarization is along the
magnetization.
}
\label{new_ishe}
\end{figure}

Consider the first case where the spin current only has $J_{33}^s$ component,
we then have
$\theta_{33k}^\mathrm{SH}=\theta_1^\mathrm{SH}M_l\epsilon_{3ln}\epsilon_{3nk}$,
which results in two possible cases ($l=1$, $n=2$, $k=1$) and ($l=2$, $n=1$, $k=2$).
Then we can obtain $J_1^c=-(\frac{2e}{\hbar})\theta_1^\mathrm{SH}M_1J_{33}^s$
and $J_2^c=-(\frac{2e}{\hbar})\theta_1^\mathrm{SH}
M_2J_{33}^s$. It says that a charge current can be generated
along the magnetization perpendicular to spin flowing direction
(as well as the spin polarization), as shown in Fig. \ref{new_ishe}(a).

For the second case where the spin current only has $J_{31}^s$ component,
we have $\theta_{31k}^\mathrm{SH}=\theta_\mathrm{SH}\epsilon_{31k}
+\theta_1^\mathrm{SH} M_l\epsilon_{3ln}\epsilon_{1nk}$.
The resulted charge current components are $J_2^c=(\frac{2e}{\hbar})
\theta_\mathrm{SH}J_{31}^s$
and $J_3^c=(\frac{2e}{\hbar})\theta_1^\mathrm{SH}M_1J_{31}^s$ where $J_2^c$ is similar
to the usual ISHE that the charge current is perpendicular to both
spin polarization and spin flow direction
while $J_3^c$ is the new term. It says that,
due to the interaction between the spin current and magnetization,
a charge current flows along the spin flowing direction when the
magnetization is along the spin polarization direction,
as shown in Fig. \ref{new_ishe}(b).

Similar to the generalized ISHE in a magnetic material, the SHE can also
exist in a magnetic material. 
The most general linear response to $\mathbf J^c$ in terms of
possible spin current $\tensor{J}^\mathrm{s}$ is
\begin{equation}
J_{ij}^s=\frac{\hbar}{2e}\theta_{ijk}^\mathrm{SH}J_k^{c},
\nonumber
\end{equation}
where $J_k^{c}$ is the charge current density along the $k$-direction and
$\theta_{ijk}^\mathrm{SH}$ is the $ijk$-component of the general spin Hall angle
tensor $\tensor{\theta}^\mathrm{SH}$ of rank 3 that depends on $\mathbf M$.
$\tensor{\theta}^\mathrm{SH}$ is given by the same expression as that of Eq.
\eqref{gen-sh}. Without losing the generality, we let charge current
along the $z$-direction. There are two possible cases:
(1) The charge current flows along $\mathbf M$,
and (2) the charge current flows perpendicular to $\mathbf M$.

In the first case,
we have $\theta_{ij3}^\mathrm{SH}=\theta_\mathrm{SH}\epsilon_{ij3}
+\theta_1^\mathrm{SH} M_3\epsilon_{i3n}\epsilon_{jn3}$. Up to the
linear term in $\mathbf M$, the first term is the usual spin Hall angle,
and the new spin currents from the second term are
$J_{11}^s=-(\frac{\hbar}{2e})\theta_1^\mathrm{SH}M_3J_3^c$
and $J_{22}^s=-(\frac{\hbar}{2e})\theta_1^\mathrm{SH}M_3J_3^c$.
It says that, due to the interaction between the charge current and magnetization,
the generated spin current flows perpendicular to charge current and
the spin polarization is along the spin flowing direction,
as shown in Fig. \ref{new_ishe}(c).

In the second case, without losing the generality we let $\mathbf M$
along the $x$-direction.
We then have $\theta_{ij3}^\mathrm{SH}=\theta_\mathrm{SH}\epsilon_{ij3}
+\theta_1^\mathrm{SH} M_1\epsilon_{i1n}\epsilon_{jn3}$. Up to the
linear term in $\mathbf M$, the first term is the usual spin Hall angle,
and the new spin current from the second term is
$J_{31}^s=(\frac{\hbar}{2e})\theta_1^\mathrm{SH}M_1J_3^c$.
It says that, due to the interaction between the charge current and magnetization,
the charge current generates a spin current of polarization
along the magnetization and flowing direction along the charge current,
as shown in Fig. \ref{new_ishe}(d).


In the current work, these principally-allowed new SHEs and ISHEs in
magnetic materials have not been considered. It is expected that the new
effects are higher orders in comparison with the usual SHE and ISHE because
they involve both spin-orbit interaction and charge-magnon interactions.
Nevertheless, it shall be very interesting to experimentally confirm these
predictions although they must exist.

\section{Conclusions}

In conclusion, a careful analysis of the electrical signals of the st-FMR in
a HM/FM bilayer has been carried out. Both rf Oersted field and rf SOTs,
which cause the ferromagnetic resonance, are considered in the analysis.
Differ from previous studies on the st-FMR, the tensor nature of the
dynamical susceptibilities is also included. It is shown that one can
indeed use dc-voltage lineshape and the angle-dependence of dc-voltage to
actually extract spin-Hall angle of the HM besides other typical magnetic
material parameters in a traditional FMR measurement.

\begin{acknowledgments}
This work was supported by the National Natural Science Foundation of
China (Grant No. 11774296, 51571109 and 11734006), Hong Kong
RGC Grants No. 16301816, 1631518 and 16300117 as well as the National
Key R{\&}D Program of China (Grants No. 2017YFA0303202 and No. 2018YFA0306004).
HFD acknowledges the support of China Scholarship Council (CSC) and
Colorado State University (CSU) during his stay in CSU.
\end{acknowledgments}


\end{document}